\documentclass[aps,prd,onecolumn,groupedaddress]{revtex4}
\pdfoutput=1
\usepackage{graphicx}
\usepackage{amsmath,amsfonts,amssymb}
\usepackage{color}
\usepackage{float}
\begin{document}

\title{A transition between bouncing hyper-inflation to $\Lambda$CDM from diffusive scalar fields} 

\author{David Benisty$^{1,2,3}$}
\email{benidav@post.bgu.ac.il}
\author{Eduardo I. Guendelman$^{1,3,4}$}
\email{guendel@bgu.ac.il}

\affiliation{Physics Department, Ben-Gurion University of the Negev, Beer-Sheva 84105, Israel$^{1}$}
\affiliation{Fachbereich Physik der Johann Wolfgang Goethe Universit\"{a}t, Campus Riedberg, Frankfurt am Main, Germany$^{2}$}
\affiliation{Frankfurt Institute for Advanced Studies, Giersch Science Center, Campus Riedberg, Frankfurt am Main, Germany$^{3}$}
\affiliation{Bahamas Advanced Study Institute and Conferences, 4A Ocean Heights, Hill View Circle, Stella Maris, Long Island, The Behamas$^{4}$}

\begin{abstract}
We consider the history of the universe from a possible big bang or a bounce into a late period of a unified interacting dark energy - dark matter model. The model is based on the Two Measures Theories (T.M.T.) which introduces a metric independent volume element and this allows us to construct a unification of dark energy and dark matter. A generalization of the T.M.T. gives a diffusive non-conservative stress energy momentum tensor in addition to the conserved stress energy tensor which appear in Einstein equations. These leads to a formulation of interacting DE-DM dust models in the form of a diffusive type interacting Unified Dark Energy and Dark Matter scenario. The deviation from $\Lambda$CDM is determined by the diffusion constant $C_2$. For $C_2=0$ the model is indistinguishable from $\Lambda$CDM. Numerical solutions of the theories show that in some $C_2 \neq 0$ the evolution of the early universe is governed by Stiff equation of state or the universe bounces to hyper inflation. But all of those solutions have a final transition to $\Lambda$CDM as a stable fixed point for the late universe.
\end{abstract}

\maketitle

\section{Introduction}
The best explanation for the accelerated expansion of our universe is the $\Lambda$CDM model, which claims that our universe mostly contains dark energy and dark matter \cite{SI1}\cite{SI2}. This model presents a big question - why the dark energy and dark matter densities in our universe are in the same order of magnitude? This problem is known as the coincidence problem. In order to solve this problem, many approaches emerged \cite{Amendola:1999er}\cite{Arevalo:2016epc}\cite{Bonometto:2017lhg}\cite{Dutta:2017fjw}. $\phi$CDM model suggests a diffusive exchange of energy between the dark energy and the dark matter \cite{cal1}\cite{cal2}\cite{haba1}. In a contrast to the basic $\Lambda$CDM model which presents our universe behaves as consisting of two separately conserved perfect fluids, representing Dark Energy and Dark Matter, $\phi$CDM implies that there is a generalized diffusive energy transfer between the two fluids that describe dark energy and dark matter. The starting point is that diffusion equation can be generalized into a curved space time by defining a non-conserved stress energy tensor $T^{\mu\nu}$ with a current source $j^\mu$:
\begin{equation} \label{chitensor}
\nabla_\mu T^{\mu\nu}=3\sigma j^\nu
\end{equation}
where $\sigma$ is the diffusion coefficient of the fluid.
The current $j^\mu$ is a time-like covariant conserved vector field $j^{\mu}_{;\mu}=0$ which describe the conservation of the number of particles in the system. Because the Einstein stress energy tensor is a covariant conserved $\nabla_\mu G^{\mu\nu}=0$ \cite{Gravity1}\cite{Gravity2}, the  $\phi$CDM model suggests that current source for dark energy stress energy tensor has to be canceled by the current source of dark matter stress energy tensor:
\begin{equation} \label{calogero}
	\nabla_\mu T^{\mu\nu}_{(\Lambda)}=-\nabla_\mu T^{\mu\nu}_{(Dust)}=J^\nu
\end{equation}
The dark energy stress tensor parametrized by a scalar field $T^{\mu\nu}_{(\Lambda)}=g^{\mu\nu}\phi$ (hence the name $\phi$CDM). This leads to another degree of freedom $\sigma$, which characterizes the exchange of energy between dark energy and dark matter.

The main problem with $\phi$CDM model is it's lack of an action principle. Therefore we developed from a generalization of Two Measure Theories \cite{intro1}-\cite{intro10} a "diffusive energy theory" which can produce on one hand a non-conserved stress energy tensor, we called $T^{\mu\nu}_{(\chi)}$, as in Eq. (\ref{chitensor}) and on the other hand a conserved stress energy tensor that we know from the right hand side of Einstein's equation, which we labeled as $T^{\mu\nu}_{(G)}$. 
This theory has some similarities to $\phi$CMD, but is not equivalent. Furthermore, this theory approaches much faster to $\Lambda$CDM behavior at late time expansion than the $\phi$CDM model . 

\section{Two Measures Theories and $\Lambda$CDM} 
The modified theory of gravity called Two Measure Theory implies another measure of integration in addition to the regular measure of integration in the action $ \sqrt{-g} $. This new measure is also a density and a total derivative. A simple example for constructing this measure, is by introducing 4 scalar fields $ \varphi_{a} $, where $ a=1,2,3,4 $. For this case,  the measure looks as:
\begin{equation}
\Phi=\varepsilon^{\alpha\beta\gamma\delta}\varepsilon_{abcd}\partial_{\alpha}\varphi_{a}\partial_{\beta}\varphi_{b}\partial_{\gamma}\varphi_{c}\partial_{\delta}\varphi_{d} 
\end{equation}
A complete action with both measures takes the form:
\begin{equation}\label{bTMT}
S=\int d^{4}x\Phi\mathcal{L}_{1}+\int d^{4}x\sqrt{-g}\mathcal{L}_{2}
\end{equation}
As a consequence of the variation with respect to the scalar fields $ \varphi_{a} $, under the assumption that $ \mathcal{L}_{1} $
and $ \mathcal{L}_{2} $
are independent of the scalar fields $\varphi_{a} $, we obtain that:
\begin{equation} \label{measure}
A_{a}^{\alpha}\partial_{\alpha}\mathcal{L}_{1}=0
\end{equation}
where $ A_{a}^{\alpha}=\varepsilon^{\alpha\beta\gamma\delta}\varepsilon_{abcd}\partial_{\beta}\varphi_{b}\partial_{\gamma}\varphi_{c}\partial_{\delta}\varphi_{d} $. Since $ \det[A_{a}^{\alpha}]\sim\Phi^{3} $ as one easily see then that for $ \Phi\neq0 $,(\ref{measure}) implies that $\mathcal{L}_{1}=M=Const$. These kind of contributions have been considered in the Two Measures Theories, which are of interest in connection with a unified model of dark energy and dark matter.

\subsection{Unified dark energy - dark energy}

A simple example, using this modified gravity approach is a model for unified dark energy and dark matter from one kinetic scalar field term \cite{m1}\cite{m2}\cite{m3}. The complete action is:
 \begin{equation}\label{udedm}
 \mathcal{S}=\frac{1}{16\pi G}\int{d^4x \sqrt{-g}R}  +\int{d^4x (\Phi+\sqrt{-g})\mathcal{L}(X,\phi)} 
 \end{equation}
where $\mathcal{L}(X,\phi)$ could be any order of k-essence term, and $X$ is the kinetic term of a scalar field  $ X =  -\frac{1}{2} g^{\alpha \beta}\phi_{,\alpha}\phi_{,\beta}$, which is different from the scalar field $\varphi_a$ we used for build the measure $\Phi$. Let's take only the firs order in $\mathcal{L}(X,\phi) = X$, but \cite{UDE} shows that higher order of the k-essence term do not change the results. From the variation with respect to the scalar field $\varphi_\alpha$, constraint on the actual value of the term $X = \alpha_1$ is obtained. In addition, the variation according to the scalar field $\phi$ gives a conserved current $j^\alpha_{;\alpha}=0$, which can be presented as:
 \begin{equation}\label{current1}
 j_\alpha = (\frac{\Phi}{\sqrt{-g}} + 1) \phi_{,\alpha}
 \end{equation}
Finally, we have to take the variation with respect to the metric, which gives us the stress energy tensor:
\begin{equation}\label{e1}
T_{\mu\nu} = g_{\mu\nu} X + j_{\mu}\phi_{,\nu}
 \end{equation}
The  Friedman-Lemaitre-Robertson-Walker (FLRW) ansatz is the standard model of cosmology form of the metric based on the assumption of a homogeneous and isotropic universe at any point, commonly referred to as the cosmological principle. The symmetry considerations lead to the FLRW metric
\begin{equation}\label{eq:robwalk}
ds^{2}=dt^{2}-a^{2}(t)\left[\frac{dr^{2}}{1-Kr^{2}}+r^{2}\left(d\theta^{2}+\sin^{2}\theta d\phi^{2}\right)\right].
\end{equation}
Herein, $a(t)$ defines the dimensionless cosmological expansion (scale) factor,
whereas $K$ denotes the positive, negative, or zero special curvature $K$ of the spatial slice.
For a cosmological solution, with the metric (\ref{eq:robwalk}), the variation with respect to the scalar field $\varphi_a$, which is used to build the new measure, implies a constant value for the kinetic term $\dot{\phi}^2=\alpha_1$. The covariant conservation of the current (\ref{current1}) gives:
\begin{equation}
j^0=\frac{\alpha_2}{a^3}
\end{equation}
where $\alpha_2$ in another integration constant. Considering the definition of $j^0$ we get:
 \begin{equation}\label{current12}
 \frac{\alpha_2}{a^3} = (\chi + 1) \dot\phi
 \end{equation}
The stress energy tensor gives the density and the pressure of that "unified scalar fluid". With equations (\ref{current1})-(\ref{current12}) we get: 
\begin{equation} \label{30}
 \rho_{DE}=\dot{\phi}^2=\alpha_1
 \end{equation}\begin{equation} \label{31}
 \rho_{Dust}=\frac{\sqrt{\alpha_2}\alpha_1}{a^3}
 \end{equation}
and the pressure is $p = -\rho_{DE}$. This solution represents a unified picture of DE-DM, and gives precisely $\Lambda$CDM model. From comparing to the $\Lambda$CDM solution, we can obtain how the observables values relate to the constant of integration that come from the solution of the theory:
 \begin{equation}\label{32}
\Omega_\Lambda=\frac{\alpha_1}{H_0} \quad , \quad 
\Omega_m=\frac{\alpha_2\sqrt{\alpha_1}}{H_0} 
\end{equation}
where $H_0$ is Hubble constant in the late universe. A generalization of this approach gives the diffusive energy action.

\section{A diffusion from dynamical time theories}
\subsection{Dynamical time action}
The constraint on the term in the action $\mathcal{L}_2$ as in the Two Measure Theories (\ref{bTMT}) can be generalized to a covariant conservation of a stress energy momentum tensor $T_{\left(\chi\right)}^{\mu\nu}$ which coupled directly in the action:
\begin{equation} \label{1}
\mathcal{S}= \int d^{4}x\sqrt{-g}\chi_{\mu;\nu}T_{\left(\chi\right)}^{\mu\nu}
\end{equation}
to a vector field $\chi_\mu$ with it's covariant derivatives $ \chi_{\mu;\nu}=\partial_{\nu}\chi_{\mu}-\Gamma_{\mu\nu}^{\lambda}\chi_{\lambda}$. From the variation with respect to the vector field $\chi_\mu$ gives a constraint on the conservation of the stress energy tensor $T_{\left(\chi\right)}^{\mu\nu}$. 
\begin{equation} 
\delta\chi_\mu: \nabla_\mu T_{\left(\chi\right)}^{\mu\nu}=0
\end{equation}
This is similar to the variation with respect to the scalar field $\varphi_a$ in the action (\ref{bTMT}), which gives $\partial_\alpha \mathcal{L}=0$. In contrast to this, in the action (\ref{1}) we get a covariant conservation of the energy momentum tensor $T_{\left(\chi\right)}^{\mu\nu}$ and not a derivative of a scalar. The correspondence between those theories is when $T^{\mu\nu}_{(\chi)}$ is taken to be as $T^{\mu\nu}_{(\chi)} = g^{\mu\nu} \mathcal{L}_m$. By introducing this term in the action (\ref{1}), we get:
\begin{equation}\label{CM}
\int d^{4}x\,\sqrt{-g}\chi_{\mu;\nu}T_{\left(\chi\right)}^{\mu\nu}=\int d^{4}x\,\sqrt{-g}\chi^{\lambda}_{;\lambda} \mathcal{L}_m= \int d^{4}x \,\partial_\mu (\sqrt{-g} \chi^\mu) \mathcal{L}_m = \int d^{4}x \, \Phi \mathcal{L}_m 
\end{equation}
As (\ref{bTMT}), the variation with respect to the scalar field gives again $\partial_\mu\mathcal{L}_m=0$, for dynamical time theories, the variation with respect to the dynamical time vector field gives this constraint too.

The reason why (\ref{1}) is called dynamical time theory is that the energy density $ T^{0}_{0} {\left(\chi\right)}$ is the canonically conjugated variable to $\chi^{0}$, which is what is expected from a dynamical time (here represented by the dynamical time $ \chi^{0} $). To show this, we take a look at the canonically conjugated momentum:
\begin{equation}\label{HMHM}
\pi_{\chi^0}=\frac{\partial\mathcal{L}}{\partial\dot{\chi^0}}=T^{0}_{0} (\chi)=\rho_{(\chi)}
\end{equation}
where $\dot\chi_0$  and the time derivative of the time component of the dynamical time vector field, and $\rho_{(\chi)}$ is the energy density of the original stress energy tensor. Notice that this action is like a regular contribution to a standard gravity theory except that instead of $\sqrt{-g}$, in that part of the action the measure of integration is the total derivative $\Phi=\partial_\mu (\sqrt{-g} \chi^\mu) $. These kind of contributions have been considered in the Two Measures Theories, which are of interest in connection with the Cosmological Constant Problem. This new definition of the measure seemingly is not made from the scalar fields as TMT, but this definition does not change the results. Cosmological solution for dynamical time theories are obtain in \cite{DT}\cite{RLSF}.
\subsection{Dynamical time action with diffusive source}
For break the conservation of $T^{\mu\nu}_{(\chi)}$ as in Eq. (\ref{chitensor}), the vector field $\chi_\mu$ should be coupled in a different part of the action, which is similar to a mass term:
\begin{equation} \label{nhd1}
S_{(\chi,A)}=\int d^{4}x\sqrt{-g}\chi_{\mu;\nu}T_{\left(\chi\right)}^{\mu\nu} +\frac{\sigma}{2}\int d^4x \sqrt{-g}(\chi_{\mu}+\partial_{\mu}A)^2 
\end{equation} 
where $A$ different scalar field from $\phi$. From a variation with respect to the dynamical space time vector field $\chi_{\mu}$ we obtain:
\begin{equation} \label{nhd2}
\nabla_{\nu}T_{\left(\chi\right)}^{\mu\nu}=\sigma(\chi^{\mu}+\partial^{\mu}A)= f^\mu,
\end{equation} 
a current source $f^\mu=\sigma (\chi^{\mu}+\partial^{\mu}A)$ for the stress energy momentum tensor $T_{\left(\chi\right)}^{\mu\nu}$.
From the variation with respect to the new scalar $A$, a covariant conservation of the current is indeed emerges: 
\begin{equation}\label{nhd3}
\nabla_{\mu}f^\mu=\nabla_{\mu}(\chi^{\mu}+\partial^{\mu}A)=0
\end{equation} 
The stress energy tensor $T_{\left(\chi\right)}^{\mu\nu}$ is substantially different from stress energy tensor that we all know from Einstein equation, which is defined as $\frac{8\pi G}{c^4}T^{\mu\nu}_{(G)}=R^{\mu\nu}-\frac{1}{2}g^{\mu\nu}R$. In this case, the stress energy momentum tensor $T^{\mu\nu}_{(\chi)}$ is a diffusive non conservative stress energy tensor. However, from a variation with respect to the metric, we get the conserved stress energy tensor as in Einstein equation:
\begin{equation}
T^{\mu\nu}_{(G)}=\frac{-2}{\sqrt{-g}}\frac{\delta(\sqrt{-g}\mathcal{L}_M)}{\delta g^{\mu\nu}}\quad , \quad \nabla_{\mu}T_{\left(G\right)}^{\mu\nu}=0 
\end{equation}
Using different expressions for $T^{\mu\nu}_{(\chi)}$ which  depends on different variables, will give the conditions between the dynamical space time vector field $\chi_\mu$ and the other variables.
\subsection{Higher derivatives action}
A particular case of diffusive energy theories is obtained when $\sigma \to \infty$. In this case, the contribution of the current $f_\mu$ in the equations of motion goes to zero, and from this constraint the vector field becomes to a gradient of the scalar:
\begin{equation}
f_\mu=\sigma(\chi_{\mu}+\partial_{\mu}A) =0 \quad \Rightarrow \quad \chi_{\mu}=-\partial_{\mu}A
\end{equation}
The theory (\ref{nhd1}) changes to a theory with higher derivatives:
\begin{equation}\label{action2}
\mathcal{S}= - \int d^{4}x\sqrt{-g} \, \nabla_\mu \nabla_\nu A \,   T_{\left(\chi\right)}^{\mu\nu}
\end{equation}
The variation with respect to the scalar $A$ gives $\nabla_\mu \nabla_\nu T^{\mu\nu}_{(\chi)}=0$ which is corresponds to the variations (\ref{nhd2}-\ref{nhd3}). In the dynamical space theory we obtain 4 equations of motion from the variation of $\chi_\mu$, which corresponds to a covariant conservation of energy momentum tensor $\nabla_\mu T^{\mu\nu}_{(\chi)}=0$. By changing the generic 4 vector to a gradient of a scalar $\partial_{\mu}\chi$ at the end, the number of conditions reduces from 4 to 1 and instead of the conservation of energy momentum tensor, we just left with a covariant conservation of the current $f^{\nu}=\nabla_\mu T^{\mu\nu}_{(\chi)}$.
\section{Scalar field Gravity with Diffusive behavior}
Our starting point is the following gravity-scalar-field action, which produces a diffusive type of interaction between dark energy and dark matter  \cite{David1}-\cite{David2} which a unification feature between these two components:
\begin{equation}\label{action}
\mathcal{S}=\frac{1}{16\pi G}\int d^{4}x\sqrt{-g}R+\int d^4x \sqrt{-g}\Lambda(\phi,X)+\int d^{4}x\sqrt{-g}\chi_{\mu;\nu}T_{\left(\chi\right)}^{\mu\nu}+\frac{\sigma}{2}\sqrt{-g}(\chi_{\mu}+\partial_{\mu}A)^2
\end{equation} 
where $\Lambda(\phi,X)$ a simple kinetic scalar field $\Lambda=-\frac{1}{2}\partial_\mu\phi\partial^\mu\phi$. For the ansatz of the stress energy tensor in the action we use a simple form which is proportional to the metric:
\begin{equation} \label{m}
T_{\left(\chi\right)}^{\mu\nu}=g^{\mu\nu}\Lambda(\phi,X) \quad \Rightarrow \quad \mathcal{S}_{(\chi)}=\int d^{4}x \sqrt{-g} \chi^\lambda_{;\lambda } \Lambda
\end{equation}
This ansatz corresponds to Eq. (\ref{CM}) where the modified measure of integration is obtained. In the case of higher derivatives theory (where $\sigma \to \infty$, and $\chi_\mu$ is replaced by a gradient of a scalar) we obtain the Galilean measure. Uses of a "Galilean measure" in String theory describe in \cite{Vulfs:2017ntw}-\cite{Vulfs:2017dhl}. As we will see, this last action will produce a diffusive interaction between DE-DM type theory. 

From the variation with respect to the vector field $\chi_\mu$ and the scalar field $A$ and we get: $\Box\Lambda=0$, whose solution will be interpreted as a dynamically generated cosmological constant with diffusion interaction. Let's undertake the important analysis of the diffusion model under the assumption of spacial homogeneity and isotropy, i.e. a space time with Friedman Robertson Walker Metric:
\begin{equation} \label{FRWM}
ds^2=-dt^2+a^2(t)[\frac{dr^2}{1-kr^2}+r^2d\Omega^2]
\end{equation}
The kinetic term becomes $\Lambda=-\partial_\mu\phi\partial^\mu\phi=\dot\phi^2$. Notice that there are higher derivative equations, but all such type of equations, correspond to conservation laws. We obtain that the variation of the scalar field (\ref{m}) will give $\frac{d}{dt}(2\dot{\phi}\ddot{\phi}a^3)=0$, which can be integrated to:
\begin{equation} \label{C2}
2\dot{\phi}\ddot{\phi}=\frac{C_2}{a^{3}} \Rightarrow \dot{\phi}^2 = C_1+C_2 \int\frac{dt}{a^{3}}
\end{equation}
From the variation with respect to the scalar field $\phi$, we get a conserved current $j^\mu_{;\mu}=0$:
\begin{equation} \label{current}
j_\alpha=2( \chi^\lambda_{;\lambda }+1)\phi_{,\alpha}
\end{equation}
For a cosmological solution we take into account only change as function of time $\phi=\phi(t)$. From that we get that the 'time' component of the current $j_\alpha$ is non zero. The conserved gives the relation:
\begin{equation} \label{25}
2\dot{\phi}(\chi^\lambda_{;\lambda }+1)=\frac{C_3}{a^3}
\end{equation}
which can be also integrated to give:
\begin{equation} \label{chi}
\chi_0=\frac{1}{a^3}\int{a^3 dt}+\frac{C_4}{a^3}-\frac{C_3}{2a^3}\int{\frac{dt}{\dot\phi}}
\end{equation}
which provides the solution for the scalar field $\chi_0$.
The last variation we should take is with respect to the metric, which gives us the (gravitational) conserved stress energy tensor: 
\begin{equation} \label{metric}
T^{\mu\nu}_{(G)}=g^{\mu\nu}(\Lambda+\chi^{\lambda}\Lambda_{,\lambda}+\frac{1}{2\sigma}\Lambda^{,\lambda}\Lambda_{,\lambda})-j^\mu\phi^{,\nu} +\chi^{\mu}\Lambda^{,\nu}+\chi^{\nu}\Lambda^{,\mu}-\frac{1}{\sigma}\Lambda^{,\mu}\Lambda^{,\mu}
\end{equation}
For cosmological solutions, the interpretation for dark energy is for term proportional to the metric $\Lambda+\chi^{\lambda}\Lambda_{,\lambda}-\frac{1}{2\sigma}\Lambda^{,\lambda}\Lambda_{,\lambda}$, and dark matter dust from the energy momentum tensor that only has a '00' is the cosmological case the tensor $-j^\mu\phi^{,\nu}+\chi^{\mu}\Lambda^{,\nu}+\chi^{\nu}\Lambda^{,\mu}+\frac{1}{\sigma}\Lambda^{,\mu}\Lambda^{,\mu}$. 
For FLRW metric we get the final terms for DE-DM densities: 
\begin{equation} \label{rhode}
\rho_{de}=\dot\phi^2+\frac{\chi_0 C_2}{a^{3}}+\frac{1}{2\sigma}\frac{C_2}{a^6}
\end{equation}\begin{equation} \label{rhodm}
\rho_{dm}=\frac{C_3}{a^3}\dot\phi-\frac{2\chi_0 C_2}{a^{3}}-\frac{1}{\sigma}\frac{C_2}{a^6}
\end{equation}
The pressure of the DE: $p_{de}=-\rho_{de}$ and for the DM: $p_{dm}=0$. This leads to the Friedman equations with (\ref{rhode})(\ref{rhodm}). For simplicity we take the limit $\sigma \to \infty$. Notice that in general the $\frac{1}{\sigma}$ terms don't contribute in the late universe. However for the early universe they may be important.

A special case is obtained when $C_2=0$, which means that the dark energy of this universe is constant $\dot{\phi}^2=C_1$. The equation of motions for the dark energy and dust  (\ref{rhode})(\ref{rhodm}) are independent on the scalar field $\chi$, and therefore the density of dust is that universe is  $\frac{C_3 \sqrt{C_1}}{a^3}$. This solution present unified non-interactive dark energy and dark matter, which corresponds to (\ref{udedm}) and its solutions (\ref{30})-(\ref{32}). This is the standard $\Lambda$CDM model with no diffusion interaction. 

When the source for the stress energy tensor $T^{\mu\nu}_{(\chi)}$ (the current $f^\mu$) is zero both stress energy tensors are conserved. The observables values related to the constant of integration that come from the solution of the theory:
 \begin{equation}
\Omega_\Lambda=\frac{C_1}{H} \quad , \quad 
\Omega_m=\frac{C_1\sqrt{C_3}}{H} 
 \end{equation}
 where $H$  is Hubble constant for the late universe. For exploring the non-trivial diffusive effect for $C_2\neq0$, we find the asymptotic solution with use of the dynamical system method. The value of $C_4$ is not affected by the evolution of the universe, because it does not appear in the density equation (\ref{rhode})-(\ref{rhodm}) where $C_2=0$. The value of $C_1=0.68$ is the ratio of dark energy $\Omega_\Lambda$, and $C_3=0.327$ is the fraction between the ratio of dark matter and the square of the ratio of dark energy $\frac{\Omega_m}{\sqrt{\Omega_\Lambda}}$.
\section{Solutions for the theory}
\subsection{Asymptotic solution and $\Lambda$CDM as a fixed point}
In order to show the stability of the solutions we use the dynamical stream line solutions. First, we normalized all the constants of integrations which appears in (\ref{rhode}-\ref{rhodm}). Basically, the dimensions of the constant of integration depend on dimensions of density and time. Therefore for the normalization we could used the critical density $\rho_c = \frac{3H^2}{8 \pi G}$ and the present value of Hubble constant $H_0$. The normalized constants are then:
\begin{subequations}
\begin{equation}
C_1:= \frac{C_1}{\rho_c} \quad , \quad C_2:= \frac{C_2}{H_0 \rho_c}
\end{equation}
\begin{equation}
C_3:= \frac{C_3}{\rho_c^{1/2}} \quad , \quad C_4:= H_0 C_4
\end{equation}
\end{subequations}
And therefore the Friedmann equation get the form:
\begin{equation}\label{DT1}
{H}=\frac{1}{\sqrt{3}}\sqrt{\dot{\phi}^2 +\frac{C_3\dot\phi- C_2 \dot\chi}{a^3}}
\end{equation}
The second Friedmann equation will give the time derivative of Hubble constant $\dot H=-\frac{2}{3}\rho_m$. From this equation we can isolate the value of the volume $a^3$. Defining a conformal time $\tau=\log(a)$ and using Friedmann equations we will give the following equations for the solutions of the theory:
\begin{subequations}
\begin{equation}\label{DTM1}
\frac{1}{H}\frac{d}{d\tau}\dot{\phi}^2=C_2 \frac{3H^2-\dot{\phi}^2}{C_3\dot\phi-C_2\chi_0}
\end{equation}
\begin{equation}\label{DTM2}
\frac{1}{H}\frac{d}{d\tau}\chi_0=1-3H\chi_0+\frac{C_3}{2\dot\phi}\frac{3H^2-\dot{\phi}^2}{C_3\dot\phi-C_2\chi_0}
\end{equation}
\begin{equation}\label{DTM3}
\frac{1}{H}\frac{d}{d\tau}H=-\frac{3}{2}(3H^2-\dot\phi^2) \frac{C_3\dot\phi-2C_2\chi_0}{C_3\dot\phi-C_2\chi_0}
\end{equation}
\end{subequations}
The coordinates of the system are ($\dot{\phi},{\chi_0},H$). An asymptotic and stable solution is obtained when the derivatives are equal to zero. Equation (\ref{DTM1}) becomes zero only for $3H^2=\dot{\phi}^2$ which means $H=\pm \sqrt{3}\dot\phi$. This equivalence also causes (\ref{DTM3}) to become zero as the first one. We obtain that the fraction $\frac{H^2-\dot{\phi}^2}{C_3\dot\phi-C_2\chi_0}$, and for (\ref{DTM2}) will become zero, the value of the scalar field $\chi_0$ has to be
\begin{equation} \label{avchi}
\chi_0=\frac{1}{3H_0}
\end{equation}
\begin{figure}[tbp]
\centering 
\includegraphics[width=1\textwidth,origin=c,angle=0]{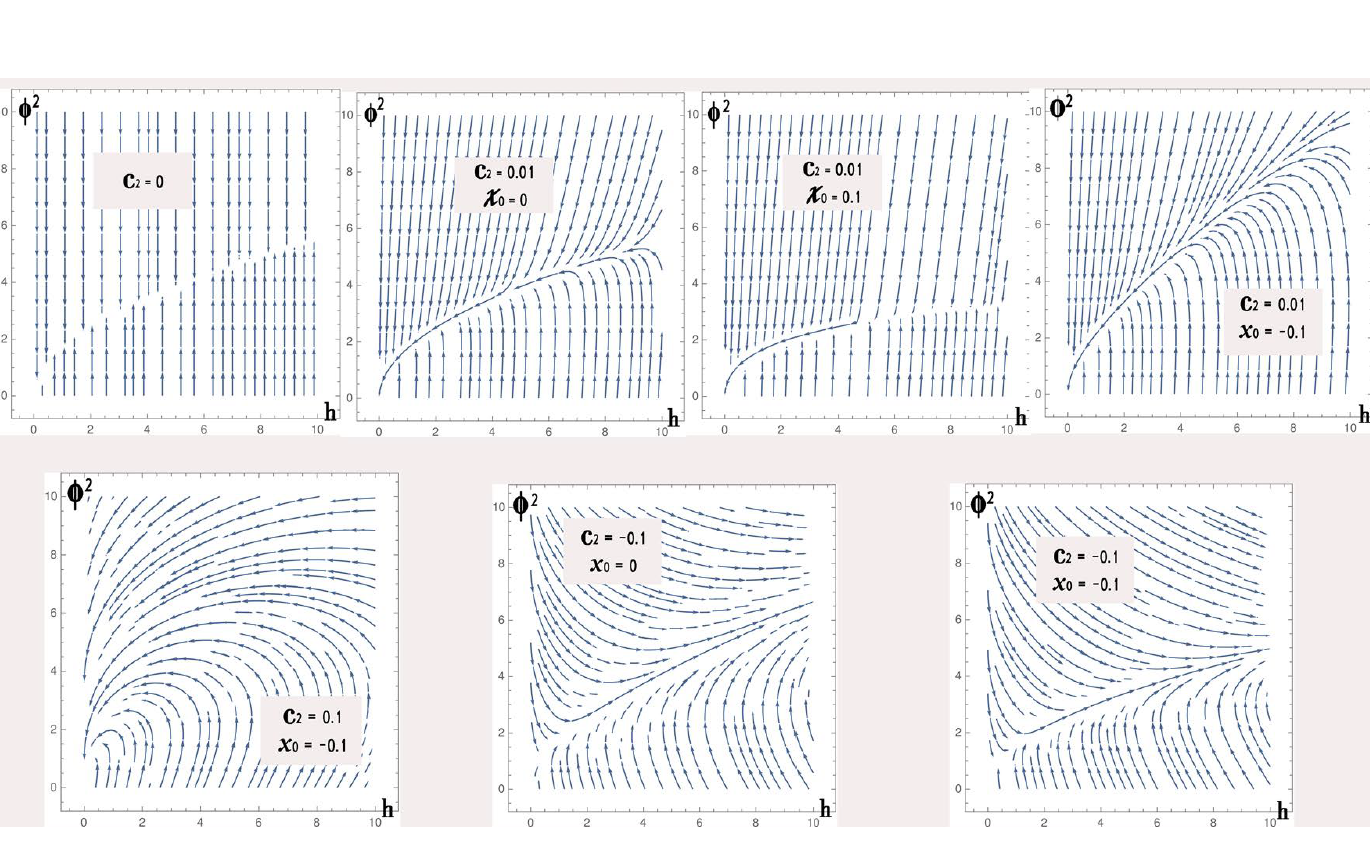}
\caption{\label{fig:4} Stream line of the dynamical variables $\dot{\phi}$ vs. $H$, for different values of $C_2$ and $C_4$. For $C_2>0$ the system collapses and for $C_2<0$ the system have an asymptotic stable solution $3H^2=\dot\phi^2$.}
\end{figure}
The asymptotic solution is given also by assuming de-Sitter space asymptoticity. First we solve for the dynamical vector field $\chi_0$ (\ref{chi}). We see that the leading term is the fraction $\frac{1}{a^3}\int{a^3 dt}$. For the asymptotic solution, where the scale factor is approximately $a(t) \approx a_0 \exp{(H_0t)}$, we obtain that there is a unique asymptotic value $\lim_{t \to \infty} \chi_0=\frac{1}{3H_0}$ as (\ref{avchi}). This is in accordance with our expectations that the expansion of the universe will stabilize the solutions into separately conserved dark energy and dark matter.

With this stability we can estimate what are the asymptotic value of dark energy and dark matter densities, from (\ref{C2})-(\ref{rhodm}). We see that in this limit, the non-constant part of $\dot{\phi}^2$ is canceled by $\frac{\chi_0 C_2}{a^{3}}$, and then:
\begin{equation} \label{arhode}
\rho_{de}=C_1+ O(\frac{1}{a^6})
\end{equation}
\begin{equation} \label{arhodm}
\rho_{dm}=(C_3\sqrt{C_1}-\frac{2C_2}{3H_0})\frac{1}{a^3}+ O(\frac{1}{a^6})
\end{equation}

The Friedmann equations provide a relation between $C_1$ and the asymptotic value of Hubble constant $H_0$ which is $H_0^2=\frac{8\pi G}{3}C_1$. For negative $C_2$ we have decaying dark energy and the last term of the contribution for dark energy density is positive (vice versa). This behavior, has a chance of explaining the coincidence problem, because unlike the standard $\Lambda$CDM model, where the dark energy is exactly constant, and the dark matter decreases like $a^{-3}$, in the case of diffusive scalar field, dark energy can slowly decrease instead of being constant and dark matter also decreases, but not as fast as $a^{-3}$.

As advanced, this behavior can be understood by the observation that in an expanding universe a non-covariant conservation of an energy momentum tensor, which may imply that some energy density is increasing in the locally inertial frame, does not mean a corresponding increase of the energy density in the co-moving cosmological frame, here in particular the non-covariant conservation of the dust component of the universe will produce  still a decreasing dust density, although for $C_2 < 0$, there will be a positive flow of energy in the inertial frame to the dust component, but the result of this flow of energy in the local inertial frame will be just that the dust energy density will decrease a bit slower that the conventional dust (but still decreases).

This is yet another example where potential instabilities are softened or in this case eliminated by the expansion of the universe. As it is known in the case of the Jeans Gravitational instability which is much softer in the expanding universe and also in other situations as well \cite{jeans}.
Another application for this mechanism could be to use it to explain the particle production from vacuum energy as expected from the inflation reheating epoch. As we see, the expansion of the universe stabilizes the solutions, such that for large times all of them become indistinguishable to $\Lambda$CDM, which appears as an attractor fixed point of our theory, showing a basic stability of the solutions at large times. 

Choosing $C_1$ as positive is necessary, because of the demand that the terms with $\sqrt{C_1}$ won't be imaginary. But for the other constants of integration, there is only the condition $C_3\sqrt{C_1}>\frac{2C_2}{3H_0} $, which gives a positive dust density at large times.  

\subsection{Stream lines and the asymptotic solution}
\begin{figure}[tbp]
\centering 
\includegraphics[width=1\textwidth,origin=c,angle=0]{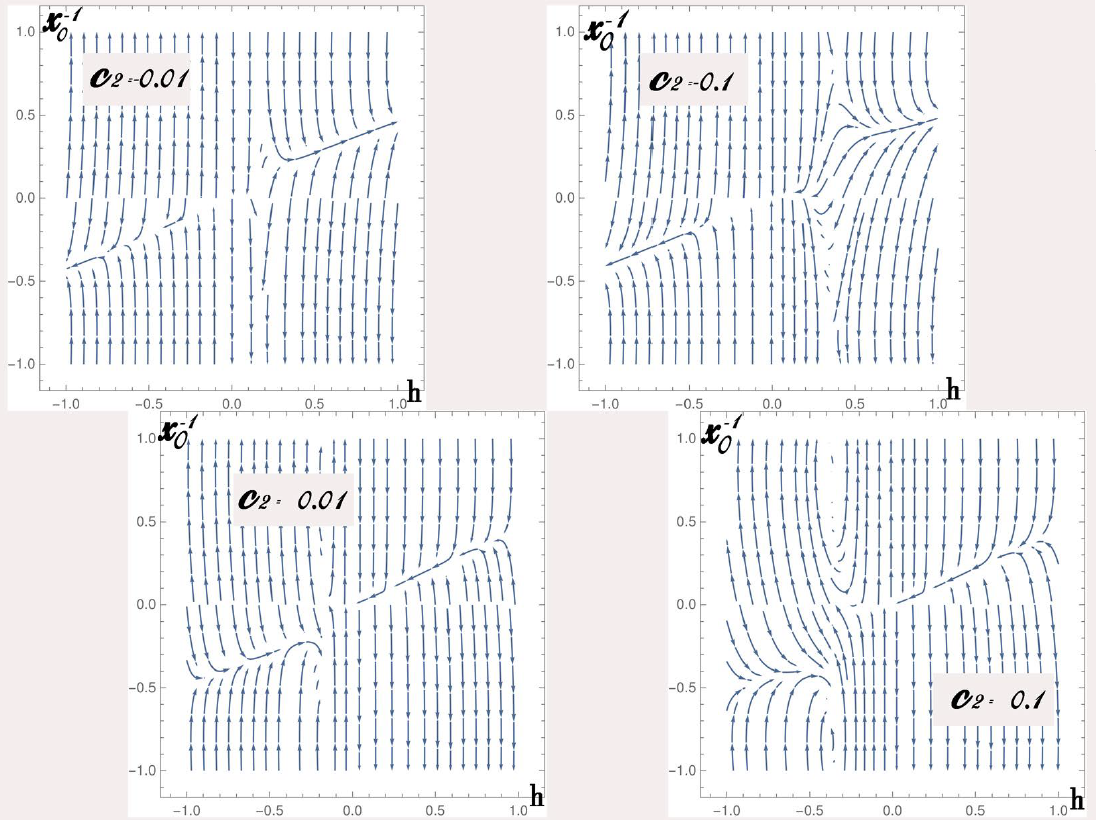}
\caption{\label{fig:5} Stream line of the dynamical variables $\chi_0^{-1}$ vs. $H$, for different values of $C_2$.}
\end{figure} 
For proving the stability of the solution and the existence of the asymptotic solutions, we plot stream lines for the dynamical system equations (\ref{DTM1}-\ref{DTM3}). Because the system is in 3D, we plot separately a couple of parameters, for any figure. In figure \ref{fig:4} we see that for positive values of $C_2$ the stream line decays to the zero point $(\dot\phi=0,H=0)$, which implies on a big crunch. In a contrast, when the diffusion constant $C_2$ have negative values, asymptoticly the stream lines look as $\dot\phi^2=3H^2$, which gives a convergence to De-Sitter space.

In figure \ref{fig:5} we see stream lines for $\chi_0^{-1}$ vs. $H$, for $\dot\phi^2=0.68$ (the asymptotic value of dark energy ratio $\Omega_\Lambda$ as \ref{arhode}). And again, for all the cases the stream lines asymptotically go to a linear connection between $\chi_0^{-1}$ and $H$ as (\ref{avchi}). But for the cases $C_2>0$ the values are decays, and for $C_2<0$ the stream lines evaluate to De-Sitter space.
With the dynamical system method it is possible to find the fixed point of the solutions of the theory. After discussing about the stream lines of the solutions, we will find numerical solutions for the theory.

\section{A transition between a bouncing hyper-inflation to $\Lambda$CDM}

For the remaining solutions, we assume that $a(t)$ is decreasing toward the past and the time variable can be replaced by the cosmological redshift variable:
\begin{equation} 
a(t) = \frac{a_0}{z(t)+1}
\end{equation}

In the case of a bounce, the assumption of $a(t)$ being decreasing toward the past breaks down. The time derivative is replaced too:
\begin{equation} 
dt = -\frac{a}{a_0}\frac{1}{H(z)}dz
\end{equation}
The equations (\ref{C2}), (\ref{chi}) can be rewriten in terms of the redshift:
\begin{equation} 
-H(z)\frac{d}{dz}\dot\phi^2=\frac{C_2}{a^2}
\end{equation}
\begin{equation} 
{H(z)}\frac{d}{dz} (a^3 \dot\chi) = -a^4 +\frac{C_3 a}{2\dot{\phi}}
\end{equation}
and the Fridmann Eq. (\ref{rhode})-(\ref{rhodm}) as well:
\begin{equation} 
{3H(z)^2}=\dot{\phi}^2 +\frac{C_3\dot\phi- C_2 \dot\chi}{a^3} - \frac{k}{a^2}
\end{equation}
where $k$ is the normalized spacial curvature of the universe. As we study from the asymptotic solution, the diffusion constant should be very small $C_2 \ll 1$ in dimensionless form. From the numerical solution we obtain that there are two different cases - where the diffusion constant is negative or positive. All of the solutions asymptotically go to $\Lambda$CDM for large times (low red shifts). In addition to the $C_2$ initial condition, we have the $C_4$ constant, which determines $\dot\chi(z=0)$. For understanding the evolution of this kind of universe, we solved numerically the deceleration rate:
\begin{equation}\label{q}
q=-1-\frac{\dot{H}}{H^2}=\frac{1}{2}(1+3\omega)(1+\frac{K}{aH^2})
\end{equation}
which shows the nature of the universe: $q=-1$ is the standard inflation ($\rho=-p$), and $q<-1$ means hyper inflation. $q=\frac{1}{2}$ is for dust dominant ($p=0$), and $q=2$ for massless scalar field, which is called Stiff equation of state ($\rho=p$).
\begin{figure}[tbp]
\centering 
\includegraphics[width=0.9\textwidth,origin=c,angle=0]{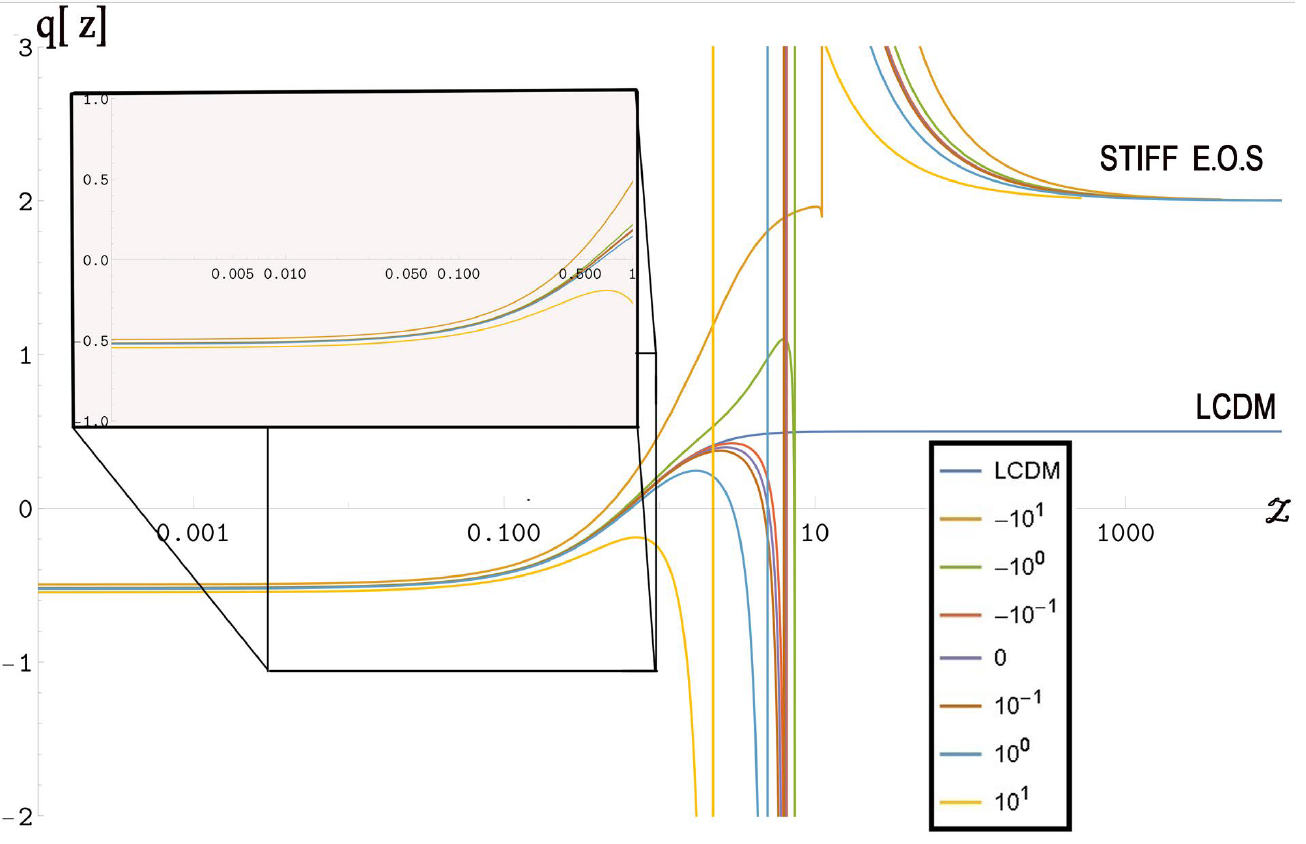}
\caption{\label{fig:1} Evolution of the deceleration parameter for different positive diffusion constant $C_2=10^{-3}$, and different values of $\dot{\chi}(0)$. We can see at that the split points of the solutions is strongly dependent on the value of $C_2$ more then $\dot{\chi}(0)$.}
\end{figure}  
\begin{figure}[tbp]
\includegraphics[width=0.9\textwidth]{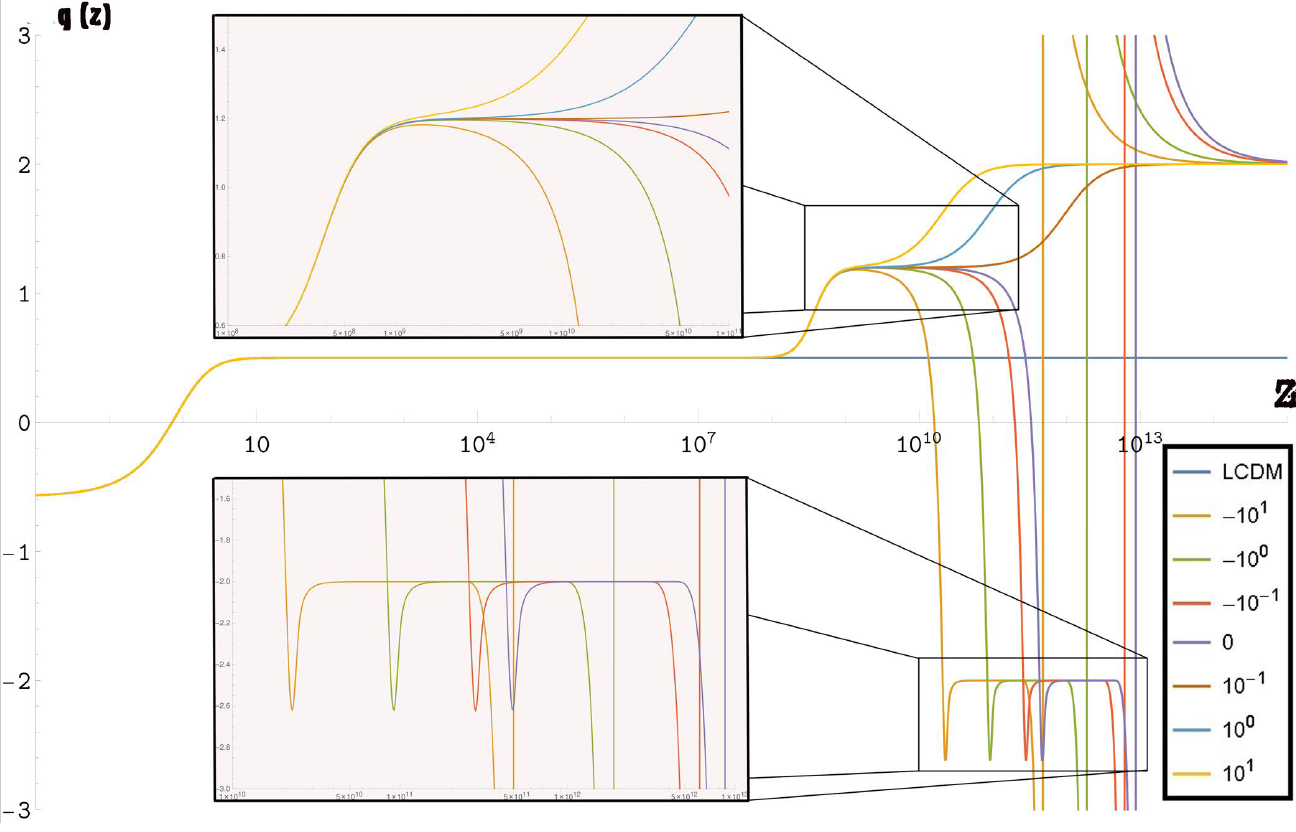}
\caption{\label{fig:2} Evolution of the deceleration parameter for negative diffusion constant $C_2=-10^{-30}$, and different values of $\dot{\chi}(0)$. In addition we can see a zoom on the point which split to different values of $\dot{\chi}(0)$, and another zoom on the hyper inflation part. The solution that approach $a=0$ are big bang - big crunch solutions, separate from the hyper inflation bouncing solution.}
\end{figure}  

\begin{figure}[tbp]
\includegraphics[width=0.9\textwidth,origin=c,angle=0]{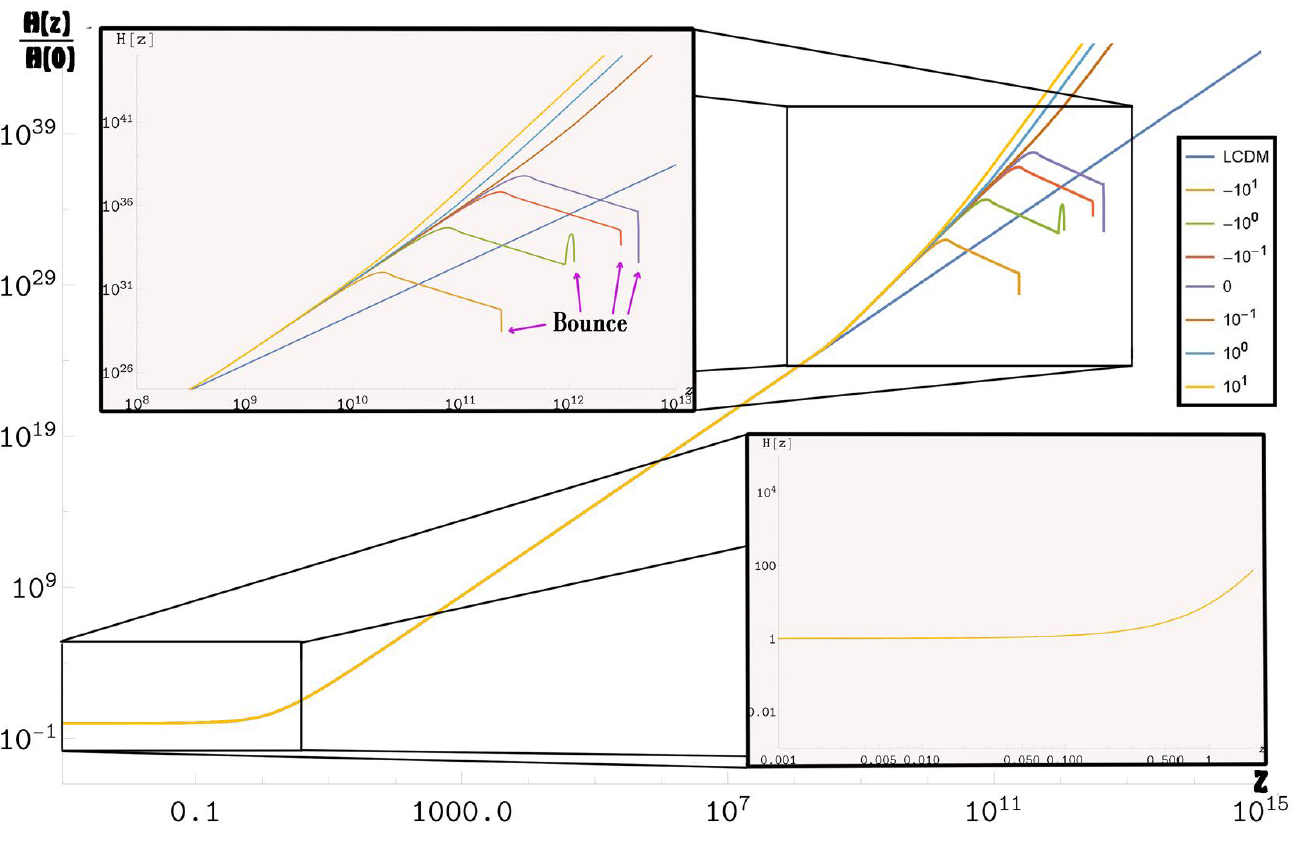}
\caption{\label{fig:3} Evolution of the Hubble constant for negative diffusion constant $C_2=-10^{-30}$, and different values of $\dot{\chi}(0)$, in a logarithmic scale. For low red shifts there isn't a change between $\Lambda$CDM and the other models. For high red shifts, in the cases of hyper inflation we can see that Hubble constant reduce to zero, which points out on a bounce.}
\end{figure}

As we can see in figure~\ref{fig:1} and figure~\ref{fig:2}, there is a different behavior for positive and negative $C_2$. For the case $C_2 > 0$, we can see a smooth change from $\Lambda$CDM at the late universe, to Stiff equation of state at the early universe directly. However, for $C_2 < 0$ we can see also the same behavior of transition, but for some values of $C_4$ we get a transition to $q=-2$ for a period of red shifts, which means hyper inflation, that from (\ref{q}) means that $\dot{H}>0$. In particular as wee see here, we can produce a bounce which means $H(z)$ goes from negative values to positive values, as in figure~\ref{fig:3}, which means $a(t)$ is not a monotonic function of time.  

A very important point is the dependence on the red shift where we can see a transition of the scalar field. The actual point of the transition can change depending on the values of $C_2$ and $C_4$ and is strongly dependent on $C_2$ then $C_4$, as one can see at figure~\ref{fig:2}. This conclusion has a big influence on the constraint on the components. From the common knowledge of the early and hot universe, there wasn't this kind of transition from hyper inflation into $\Lambda$CDM in the nucleosynthesis epoch. Therefore a big constraint on the values of $C_2$ and $C_4$ is coming from requirement that this kind of phase transition would happen only before BBN ($z=10^8$) at least.

In the case we restrict ourselves to $k=0$ the value of $C_2$ is of the order of $C_2 \preceq 10^{-15}$ and $\dot{\chi}(0)$ which affects the duration of the hyper-inflation, but there are no 60 e-folds needed to solve the flatness problem. One has to point out however, that the importance of the super acceleration period is not so much to produce many e-foldings, but to produce the bouncing of the universe (see for example \cite{b1}\cite{b2}\cite{b3}), and therefore obtain the avoidance of the initial singularity. In a future publication we will see how to extend the period of inflation to obtain 60 e-foldings. In addition to hyper inflation there are separate different big bang - big crunch solutions, where the universe starts at $a=0$ and goes back to $a=0$. We depict that prediction in figure~\ref{fig:2}. 

Even for small spacial curvature, the first possibility of bounce happens for different red shifts. Indeed, a prediction of standard inflation theory, the curvature of our universe should be a exactly zero, however from measurements there are only strict constraints on the spacial curvature. And for our model, if there is a small curvature, the red shift where the phase transition takes place changes.
\section{Discussion}
In this paper we generalize the Two Measure Theories and the dynamical space time theory, by couples the dynamical time vector field to another scalar field. We obtain a covariant conservation of a current source for diffusive energy momentum tensor which is introduced in the action. This current is dissipated in the case of an expanding universe. The stress energy tensor which is related to this current, is not the known gravitational energy tensor which appears in the right hand side of the Einstein equation. The non-covariant conservation of the energy momentum tensor that appears in the action induces an energy transfer between the dark energy and dark matter components in the gravitational energy momentum tensor. In a way that resembles the ideas in \cite{haba1}. But the $\phi$CMD model lacks from an action principle. Although the mechanism is similar the formulation are not equivalent.

From the asymptotic solution we obtain that when $C_2<0$, unlike the standard $\Lambda$CDM model (which has a constant dark energy density and the dark matter decreases like $a^{-3}$) in our case, dark energy can slowly decrease, instead of being constant, and dark matter also decreases, but not as fast as $a^{-3}$. This behavior, where $C_2 < 0$ has a chance of explaining the coincidence problem and is also different from the $\phi$CMD model, where the exchange between dark energy and dark matter is much stronger in the asymptotic limit.

This behavior is considered by the observation that in an expanding universe a non-covariant conservation of an energy momentum tensor, which may imply that some energy density is increasing in the locally inertial frame does not mean a corresponding increase of the energy density in the co-moving cosmological frame. Here in particular, the non-covariant conservation of the dust component in the energy momentum tensor still produces a decreasing dust density, although for $C_2 < 0$ a positive flow of energy in the inertial frame to the dust component exits. However, the result of this energy flow in the local inertial frame is that the dust energy density decreases a bit slower then the conventional dust.  

From the numerical solution we obtain two interesting cases: One case is when the universe starts from Stiff equation of state, and have a transition into $\Lambda$CDM behavior. In a few solutions also the universe becomes to a "dark radiation" dominant, before it evolves into $\Lambda$CDM. In another case, the bounces take place at very high red shifts if $C_2$ small enough, and therefore they have no singularities. The bounce is consequence of the hyper inflation period, where the deceleration parameter is $q=-2$, and $\dot{H}>0$ (so H goes from negative values to $H=0$ and then to a positive value). The hyper inflation itself does not give enough e-folds in the most simple version of the theory (not more than 4 e-folds), but although inflation address many questions like the horizon and the flatness problem, it does not solve the singularity problem, which the bounce solves. In the future we will try to generalize this theory, and give a more complete picture that fits for our universe. In ref. \cite{BS} a similar model has been studied. However, in \cite{BS}  $\Lambda$ is fundamental field (not defined a in terms of a scalar field $\phi$)  and dark matter - dark energy unification is not discussed, but some type of bounce behavior is also found. 

\appendix
     
\section{Non-covariantly conserved Stress energy tensor}
An equivalent expression for (\ref{chitensor}), when $T^{\mu\nu}_{(\chi)}$ is formulated as a perfect fluid in FRWM space is:
$$\dot{\rho}+3\frac{\dot{a}}{a}(\rho+p)=\frac{C_2}{a^3} $$ 

In the case the diffusion constant is $C_2=0$, the stress energy tensor is conserved, and there is no diffusive effect. For late times, where the scale parameter goes to infinity, we obtain that the diffusive effect vanishes.

\acknowledgments
This article is supported by COST Action CA15117 "Cosmology and Astrophysics Network for Theoretical Advances and Training Action" (CANTATA), of the COST (European Cooperation in Science and Technology). We thank Sebastian Bahcmonde for very useful discussions.

\end{document}